# Accurately measuring phase profiles of structured light in optical manipulation

Xionggui Tang*, Yanhua Xu, Yi Shen and Wenjing Rong

*Department of Physics, Key Laboratory of Low Dimensional Quantum Structures and Quantum Control of Ministry of Education, Hunan Normal University, Changsha, 410081, P.R. China*
*tangxg@hunnu.edu.cn

**Abstract:** Accurate phase measurement is highly desirable in optical manipulation driven by structured light, since the phase profiles have a close connection with optical force distribution, strongly affecting the capability of optical traps. Here, we report a novel interferometric phase measurement technique for accurately measuring the phase profiles of structured light in holographic optical trapping system. The related experiments are performed, in which the obtained phase profiles agree well with the designed phase profiles. It reveals that the phase profiles of structured light can be accurately obtained. Importantly, the fast switch between phase measurement and optical manipulation experiment can be easily realized, which are strongly preferable in optical trapping system. This study offers a promising way for accurately obtaining phase profiles of structure light, which is very helpful for realizing optical trapping with high performance.

## 1. Introduction

Structured light is a tailored beam, in which optical field has unique profile, indicating that the intensity, phase and polarization typically exhibit strong spatial inhomogeneity [1-3]. For instance, optical vortices typically possess helical phase profiles, carrying orbit angular momentum (OAM) per photon of $m\hbar$ ($m$ denotes topological charge and $\hbar$ is the reduced Planck constant) [4]. Experimentally, the structured lights are created by usually shaping the amplitude, phase, and polarization of a fundamental Gaussian beam via diffractive optical elements, holograms, metasurface and other tools [2]. In the past two decades, structured lights have experienced a rapid development, and now become an intense research area. Because of extraordinary optical properties of structured light, it offers new possibilities in wide variety of applications including super-resolution imaging, quantum information process, optical communications, and optical manipulation [5~7]. In these different applications, accurate generation and detection of various structured light beams is highly desirable, especially in optical systems for particle trapping and transportation. Previously, several approaches for detection of structured light have been proposed, such as mode sorters or modal decomposition for detecting optical vertices [2, 8]. For these approaches, however, the accuracy of phase measurement still remains relatively poor. Thus, these approaches aren't capable of accurately measuring light beams with complex phase profiles and large phase gradient at small footprint. In contrast, the optical force driven by phase gradient in optical manipulation systems strongly depends on phase profiles, while steering and sorting nanoparticles in holographic optical tweezers [9~11]. Accordingly, it is of significance to develop a new technique for accurately detecting phase profiles of structured light in optical trapping system.

Nowadays, phase measurement has wide applications in many fields, such as microscope, imaging through scattering media, beam shaping and characterization of optical elements [12-15]. Over past several years, different approaches for measuring phase of light have been demonstrated [16-22]. These methods can be divided into two groups. The first group is called the non-interferometric phase measurement [16-18]. The Shack-Hartmann wavefront sensor is a good example [18, 19], which consists of an array of lenses in the front of image sensor, as designed by geometrical optics. In principle, each lens enables measurement of the average phase slope in the area of lens, obtained by its location of optical spot on the image sensor. Although the improvement has been achieved recently, its weakness such as low spatial resolution still remains. Evidently, it is only used for detecting the smooth phase profiles. The second group is interferometric phase measurement, in which the phase profiles are obtained by interferometric fringes [20-22]. Although it can offer relatively high spatial resolution, there still exists certain limitations that the delicate interference system are highly sensitive to slight vibrations, air perturbation or spatial alignment, easily leading to a distinct decrease in accuracy. However, the phase profiles of structured light for optical manipulation typically have high complexity, including optical phase singularity, large phase gradient, high asymmetry, small footprint, which are strongly associated with the desired optical force maps. Indeed, these phase profiles still remains difficult to be accurately detected, and the technique for measuring these phase profiles has not been reported yet.

In this work, we demonstrate a novel interferometric phase measurement method for obtaining the phase profiles of structured light in optical manipulation, in which the two beams used for optical interference have the same intensity distribution and common propagation path. It exhibits high accuracy, strong flexibility, no environmental noise. Importantly, the switch between the optical manipulation and phase measurement can be easily and quickly implemented, which are strongly desirable for optical manipulation of small particles.

## 2. Method

Nowadays, holographic optical tweezers has become a powerful tool for optically trapping and manipulating various particles, due to its high flexibility of dynamically creating various structured light. Herein, our approach for measuring phase of structured light is directly carried out in holographic optical tweezers. The schematic diagram of the experimental setup is illustrated in Fig. 1(a). Firstly, a reflective spatial light modulator (SLM) is illuminated by a collimated laser beam. Then, its reflected beam is modulated by a designed hologram loaded on SLM, as given in Fig. 1b(i), and propagates along optical axis of 4-$f$ system (i.e. telescope system). Next, its optical field is imaged on the the back focal plane of microscope. Lastly, the optical pattern is generated at the focal plane of microscope, which is the spatial frequency spectrum of the optical field reflected by SLM. Typically, the quality of optical pattern has a strong impact on optical manipulation performance, since the intensity and phase gradient force are largely determined by the intensity and phase profiles. A beam profiler with high resolution can be employed for accurately capturing the optical intensity at the focal plane. For phase measurement, however, it still remains unsolved. Here, we propose a novel interferometric method, as called common-path interference method, in which two interference beams have not only same intensity profiles but also common propagation path. Accordingly, we just need design another hologram shown in Fig. 1b(ii) to generate reference beam at the focal plane of microscope, whose phase profile has a specific distribution. In this case, two holograms can be rapidly designed by using our method reported previously [23], and their superposition is performed as a combined hologram for rapidly and conveniently achieving optical interference pattern at detection plane, as depicted in Fig. 1b(iii). Then, the optical interference pattern can be captured by beam profiler, as illustrated in Fig. 1b(iv). Finally, the phase shifting method is employed to obtain the phase profile of the structured light [24], as shown in Fig. 1b(v). For the measured phase profile, its quality can be evaluated by using the Root Mean Squared Error (RMS), and it is written as,

$$rms = \sqrt{\frac{1}{N}\sum_{i-1}^{N}[\theta_i^a - \theta_i^b]^2} \qquad (1)$$

where $\theta_i^a$ and $\theta_i^b$ denote the actual phase profiles and the measured phase profiles at the position $(x_i, y_i)$, respectively.

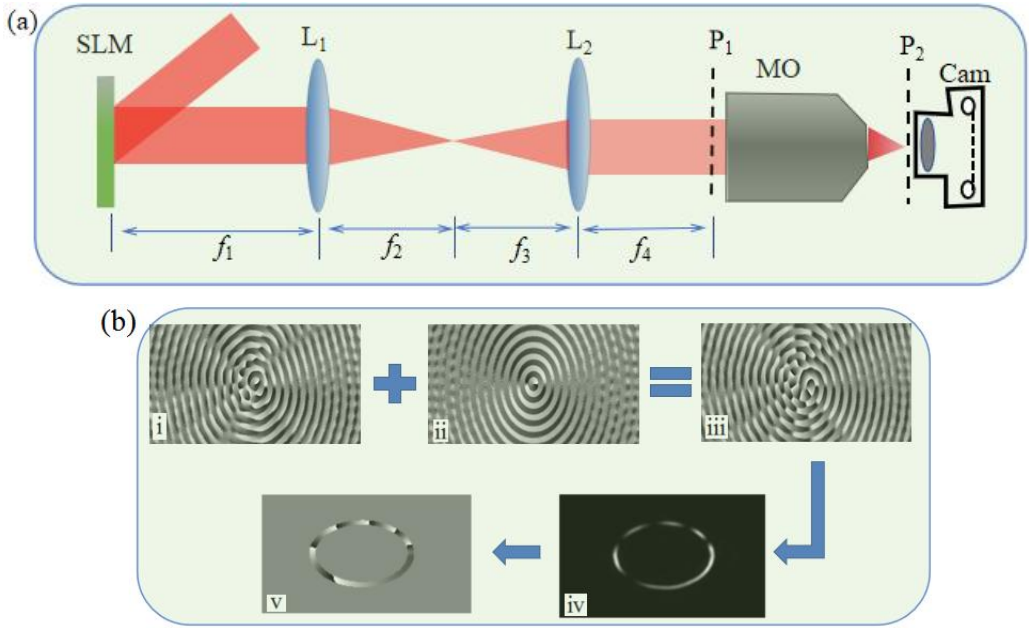

Fig. 1. The phase measurement of structured light, (a) Schematic diagram of setup, (b) Flow chart of obtaining phase profile. (SLM: Spatial light modulator; L1-L2: Lens; MO: Microscopic objective; f1-f4: focal length; P1: the back focal plane of microscope; P2: the focal plane of microscope; Cam: Camera)

While compared with the conventional phase measurement method, evidently, our scheme has several advantages including high accuracy, measurement rapidness, strong flexibility, due to fact that two interference beams are directly created by their corresponding holograms. Consequently, the phase measurement is directly performed in holographic optical trapping system, in which the holograms for creating the two beams can be easily designed. In this case, the fast switch between the optical manipulation and phase measurement can be conveniently operated, and it has no negative effects on optical manipulation of various particles.

## 3. Results and analysis

In this section, we first demonstrate the validity of our proposed method by using a simple structured light. In this case, a ring optical pattern with linear phase distribution, which is typically used for optical manipulation in optical trapping system, is chosen as an example for evaluating phase measurement. The holograms for creating ring optical pattern and its reference beam are shown in Fig. 2 (a) and (b), respectively. Then, the combined hologram is calculated, as presented in Fig. 2(c), and is directly loaded by a phase-only SLM in holographic optical trapping system. Next, it interference optical pattern is recorded, as given in Fig. 2(d). Finally, its recovered phase profile is obtained as shown in Fig. 2(e). It finds that the obtained phase profile is in good consistent with the designed one, as depicted in Fig. 2(f), revealing that the phase profile is very accurate.

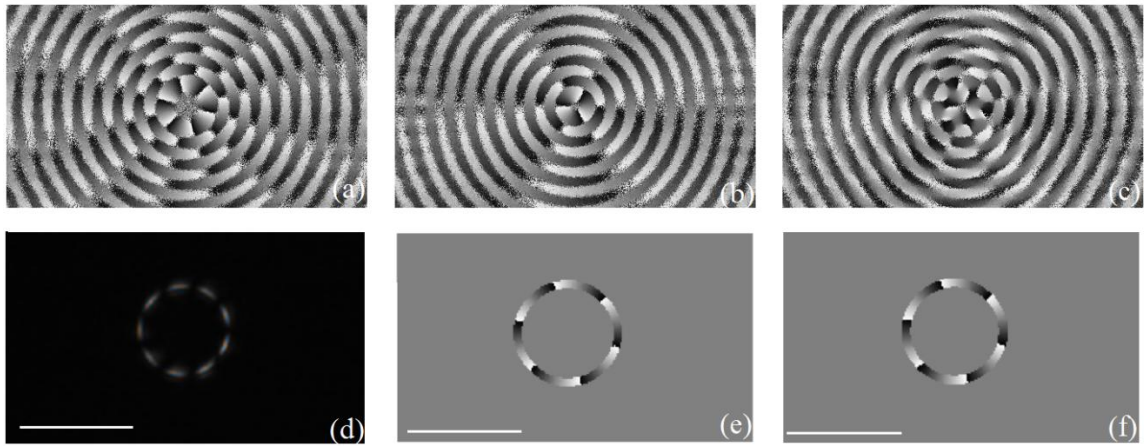

Fig. 2. (a) Hologram of ring optical pattern, (b) Hologram of reference beam, (c) The combined hologram, (d) The measured interference pattern, (e) The obtained phase profile, (f) The designed phase profile. The scale bar is 10 μm.

In the following, several different optical patterns with complex phase profiles are illustrated for further testing whether the measured phase profiles can meet our requirement. Here, the ring, oval, rectangle and star optical patterns with nonlinear phase profiles are created by the designed holograms. The corresponding designed phase profiles of optical patterns at output plane are shown in Fig. 3a(i~iv), respectively. Then, our interferometric method is carried out, and its interference patterns are recorded at output plane, as presented in Fig. 3b(i~iv), respectively. Accordingly, the recovered phase profiles are obtained by phase shifting method, as given in Fig. 3c(i~iv), respectively. Apparently, it shows that the measured phase profiles agree well with the designed ones, which also indicates that complex phase profile can be accurately obtained. Similarly, the measured phase profiles of other optical patterns also have high accuracy, which are not provided here. As a result, our proposed scheme can offer a powerful tool for phase measurement of structured light, which is useful for realizing optical manipulation with high performance.

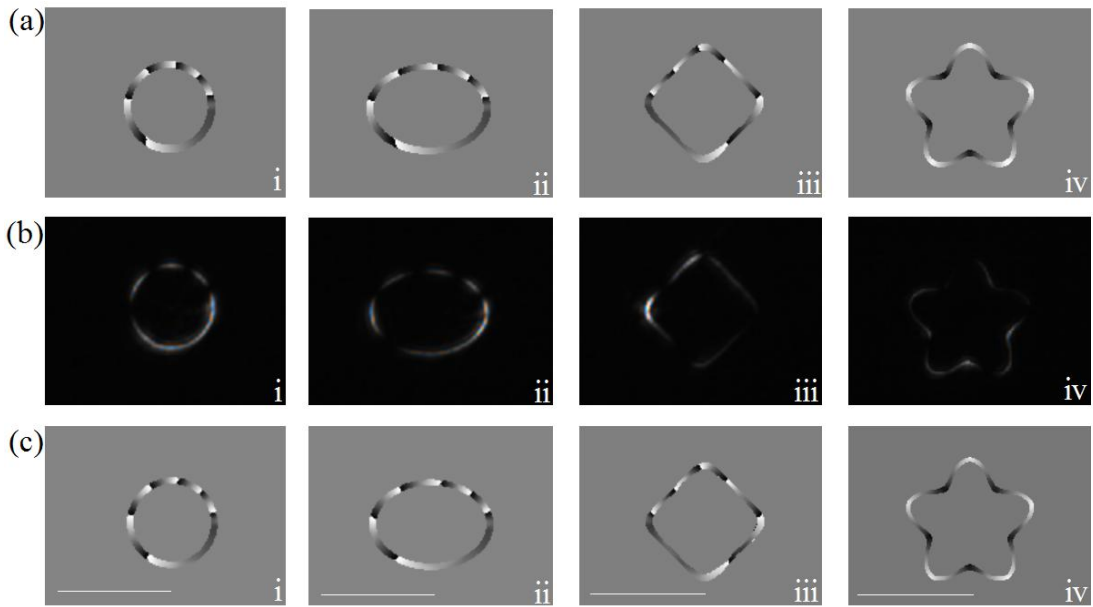

Fig. 3. Different optical pattern with nonlinear phase profile, (a) The designed phase profiles, (b) the measured interference patterns, (c) The obtained phase profiles. Panels i-iv correspond to ring, oval, rectangle and star optical patterns. The scale bar is 10 μm.

Lastly, phase profiles obtained by our proposed method are compared with ones by conventional interference method, to investigate phase accuracy difference between them. For simplicity, the ring optical pattern with different phase profiles are chosen to be demonstrated. Their designed phase profiles of ring optical patterns are presented in Fig. 4a(1-iii), which have linear, nonlinear and symmetrical distribution, respectively. Afterward, the phase profiles obtained by our proposed method are given in Fig. 4b(i-iii), and their root mean squared error are $3.16 \times 10^{-2}$ Rad, $2.64 \times 10^{-2}$ Rad and $3.31 \times 10^{-2}$ Rad, respectively, revealing that the phase accuracy are high. In contrast, the phase profiles by using conventional interference method are shown in Fig. 4c(1-iii), respectively. However, their root mean squared error are $1.31 \times 10^{-1}$ Rad, $1.34 \times 10^{-1}$ Rad and $1.50 \times 10^{-1}$ Rad, respectively. Apparently, the root mean squared error in Fig. 4(c) are much larger that those in Fig. 4(b). Additionally, the phase accuracy by conventional interference method is also affected by other factors including the amplitude and incident direction of reference beam, which requires that those parameters should be precisely controlled.

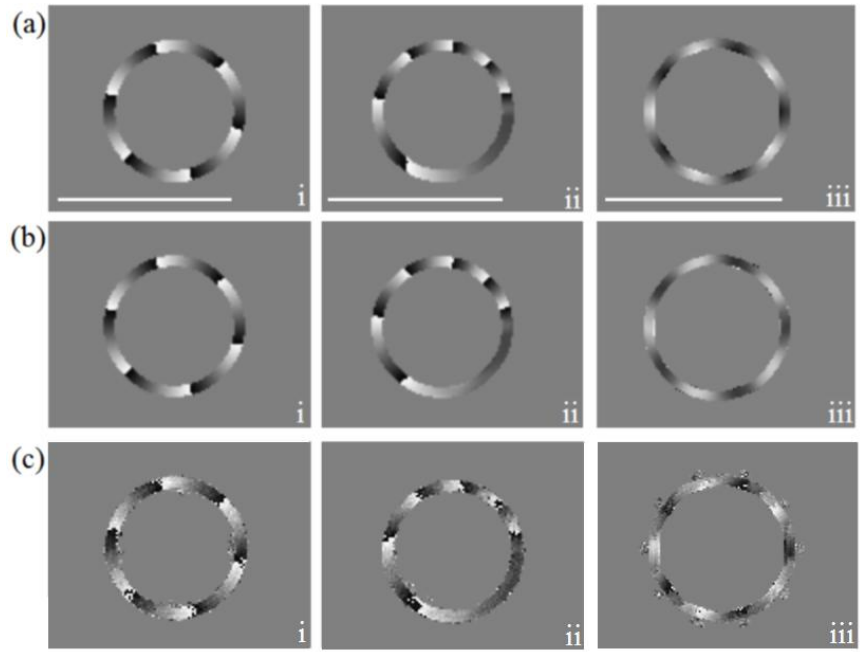

Fig. 4. (a) The designed phase profiles, (b) The phase profiles obtained by our method, (c) The phase profiles obtained by conventional interference method. Panels i-iii correspond to ring with linear, nonlinear and symmetric phase profile. The scale bar is 10 μm.

## 4 Conclusions

In this work, we have proposed and experimentally demonstrated a common-path interference method for phase measurement of structured light in optical trapping system, in which the two beam have same intensity distribution and propagation path. The experimental results show that the phase profiles by our proposed method exhibit high accuracy, but those obtained by conventional interference method apparently have poor accuracy. Importantly, our proposed method has remarkable advantages including high accuracy, strong flexibility, no noise perturbation, and measurement rapidness, which are very useful for optical manipulation of various small particles. Accordingly, our study provides a promising technique for accurate phase measurement of structure light.

**Declaration of Competing Interest**

The authors declare that they have no known competing financial interests or personal relationships that could have appeared to influence the work reported in this paper.

**Acknowledgment**

This work was supported by the Scientific Research Foundation of Hunan Provincial Education Department under Grant 20A315, and the Scientific Research Foundation of Changsha City under Grant kq2202238.